\documentclass{article}
\usepackage{ijcai25}

\usepackage{times}
\usepackage{soul}
\usepackage{url}
\usepackage[hidelinks]{hyperref}
\usepackage[utf8]{inputenc}
\usepackage[small]{caption}
\usepackage{graphicx}
\usepackage{amsmath}
\usepackage{amsthm}
\usepackage{booktabs}
\usepackage{algorithm}
\usepackage{algorithmic}
\usepackage[switch]{lineno}
\usepackage{subfigure}
\usepackage{cite}
\usepackage{multirow}
\usepackage{adjustbox}
\usepackage{makecell}
\usepackage{bm}
\usepackage{amssymb}
\usepackage{color}
\usepackage[table]{xcolor}
\usepackage{xcolor}
\usepackage{arydshln}
\usepackage{diagbox}
\usepackage{pifont}
\usepackage{natbib}
\title{Weakly-supervised Audio Temporal Forgery Localization via Progressive Audio-language Co-learning Network}

\author{
Junyan Wu$^1$
\and
Wenbo Xu$^1$
\and
Wei Lu$^{1,}$\thanks{Corresponding author.}
\and
Xiangyang Luo$^2$
\and
Rui Yang$^3$
\and
Shize Guo$^2$
\\
\affiliations
$^1$ MoE Key Laboratory of Information Technology, Sun Yat-sen University, Guangzhou, China\\
$^2$ State Key Laboratory of Mathematical Engineering and Advanced Computing, Zhengzhou, China\\
$^3$ Alibaba Group, Hangzhou, China\\
\emails
wujy298@mail2.sysu.edu.cn, xuwb25@mail2.sysu.edu.cn,
luwei3@mail.sysu.edu.cn, \\
luoxy\_ieu@sina.com,
duming.yr@alibaba-inc.com,
{guosz\_ieu}@sina.com
}
\begin{document}

\maketitle

\begin{abstract}
Audio temporal forgery localization (ATFL) aims to find the precise forgery regions of the partial spoof audio that is purposefully modified. Existing ATFL methods rely on training efficient networks using fine-grained annotations, which are obtained costly and challenging in real-world scenarios. 
To meet this challenge, in this paper, we propose a progressive audio-language co-learning network (LOCO) that adopts co-learning and self-supervision manners to prompt localization performance under weak supervision scenarios.
Specifically,  an audio-language co-learning module is first designed to capture forgery consensus features by aligning semantics from temporal and global perspectives.
In this module, forgery-aware prompts are constructed by using utterance-level annotations together with learnable prompts, which can incorporate semantic priors into temporal content features dynamically. 
In addition, a forgery localization module is applied to produce forgery proposals based on fused forgery-class activation sequences.
Finally, a progressive refinement strategy is introduced to generate pseudo frame-level labels and leverage supervised semantic contrastive learning to amplify the semantic distinction between real and fake content, thereby continuously optimizing forgery-aware features. 
Extensive experiments show that the proposed LOCO \footnote{Code and pre-trained models are available at \url{https://github.com/ItzJuny/LOCO}.} achieves SOTA performance on three public benchmarks.
\end{abstract}

\section{Introduction}

\begin{figure}[h]
    \centering
    \includegraphics[width=1\linewidth]{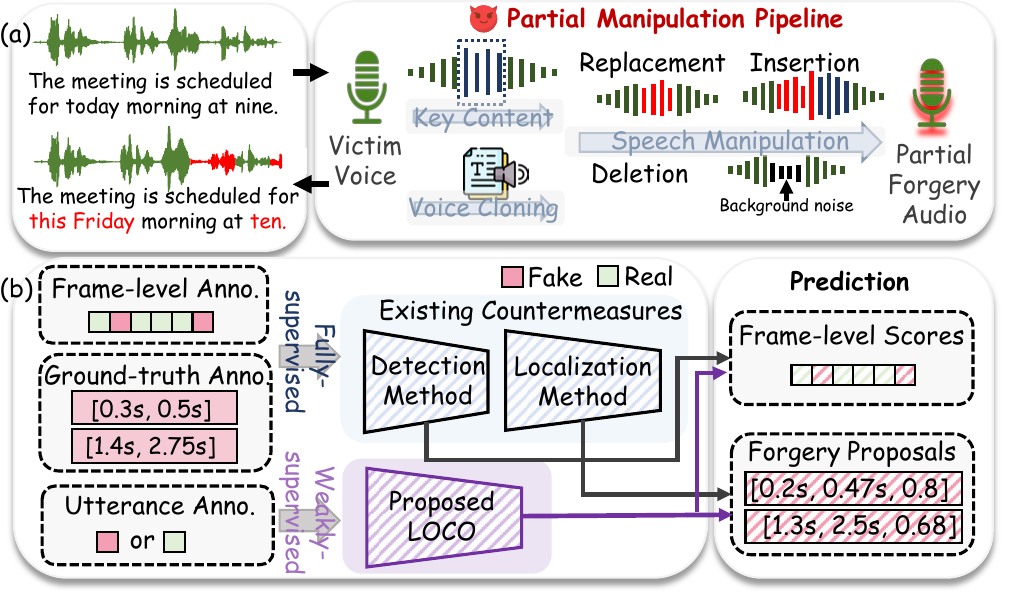}
    \caption{(a) The pipeline of partial forgery audio manipulation, which poses an urgent need to study advanced detection and localization methods. (b) Differences in training between existing countermeasures and the proposed LOCO, highlighting the challenges of weakly-supervised learning from utterance-level labels.}
    \label{fig:1}
\end{figure}
\vspace{-0.5em}

With the development of Artificial Intelligence Generated Content (AIGC), more creative and realistic productions are coming into the limelight \cite{tts1, tts2, groot, l1,l2,l3}. 
Recently, the malicious use of AIGC for low-cost audio partial forgery manipulation (PFM) has raised people's safety concerns \cite{AVDF1M}. 
As shown in Figure \ref{fig:1} (a), a malicious attacker manipulates small portions of the real audio to alter the original meaning. It can easily evade the detection of utterance-level countermeasures \cite{AASIST, SSLAS, AMSDF} since the difference between these audio is small.
To defend against PFMs, several datasets and advanced countermeasures \cite{IFBDN, TDL-ADD} are proposed. Despite the remarkable progress achieved, some issues still exist.

(1) \textbf{Training Manner}: Existing PFM countermeasures adopt a fully-supervised training manner \cite{CFPRF, BATFD, BATFD2}, which requires fine-grained annotations, such as frame-level labels and ground-truth annotations depicted in Figure \ref{fig:1}(b). However, in real-world scenarios, fine-grained annotations are costly and difficult to obtain. In this case, research on weakly-supervised temporal forgery localization (wATFL) is more popular and practical, yet such approaches remain underdeveloped.

(2) \textbf{Weak Supervision Challenge}:  Under a weakly-supervised training manner, it is highly challenging to locate subtle forgery regions based solely on utterance-level (i.e., binary) labels. Therefore, it is vital to develop countermeasures guided by the intrinsic principles of PFMs to prompt localization performance in weakly-supervised scenarios. 



In this paper, we are motivated to introduce a weakly supervised learning \underline{lo}calization method, namely progressive audio-language \underline{co}-learning network (LOCO), which, to the best of our knowledge, is being explored for the first time in weakly-supervised partial forgery audio localization.
Specifically, we first design the audio-language co-learning (A2LC) module to capture forgery consensus features by aligning semantics from temporal and global perspectives. In this module, a temporal forgery attention (TFA) adapter is used to model temporal forgery cues for semantic features.
In addition, a prompt-enhanced forgery feature (PFF) adapter is devised to dynamically incorporate forgery-aware prompts into content semantics with valuable contextual and semantic cues from a global view. 
Through a co-learning manner, forgery-aware consensus features can be learned to capture semantic inconsistency cues raised from PFM. 
Moreover, due to the lack of fine-grained annotations, it is intuitive to obtain frame-level labels for free from a self-supervised perspective.
Thus, we devise a progressive refinement strategy (PRS) that gradually guides the optimization of forgery-aware features through supervised semantic contrastive learning (SCL) to further amplify the semantic distinction between real and fake content.


To sum up, the main contributions of this paper are presented as follows:


\begin{itemize}
\item We propose a novel wATFL method named LOCO, which promotes the learning of forgery-aware features in precise detection and localization performance through co-learning and self-learning manners.
\item We specifically design an A2LC module to mine the semantic inconsistencies caused by PFM and introduce PRS to progressively amplify the semantic differences between real and forgery content. 
\item We conduct extensive experiments to demonstrate that the proposed LOCO performs favourably against the state-of-the-art methods on three public datasets.
\end{itemize}


\section{Related Work}
\subsection{Temporal Forgery Localization}
Temporal forgery localization aims to determine the precise timestamps of forgery manipulation segments within an audio or video. 
To advance the research, various datasets \cite{HAD, PSDL, BATFD, AVDF1M} have been proposed and can be divided into two categories based on whether semantic information is modified. 


BA-TFD \cite{BATFD} was the first framework designed for the temporal forgery localization task, and it adopts a 3D-CNN backbone with a boundary matching network (BMN) for localization, where features are guided by contrastive learning, boundary matching, and frame classification. 
Subsequently, BA-TFD+ \cite{BATFD2} replaced the backbone with a multi-scale transformer and enhanced the localization module with a BSN++ network. 
Additionally, UMMAFormer \cite{UMMAF} introduced a novel temporal feature abnormal attention module, and a parallel cross-attention feature pyramid network to enhance features from sequential multimedia data and adopted an ActionFormer decoder for localization.
Moreover, CFPRF \cite{CFPRF} devised a frame-level detection network in the first stage to learn robust representations for better indicating rough forgery regions and employed a proposal refinement network in the second stage to produce fine-grained proposals.

Despite the success of fully-supervised ATFL methods, they require massive and consuming frame-level artificial annotations, which limits their practicality in real-world scenarios where fine-grained annotations are unavailable.
\vspace{-0.2em}
\subsection{Weakly-supervised Learning}
Currently, wATFL methods have not been explored, with the closest being weakly-supervised temporal action localization (wTAL) and video anomaly detection (wVAD). Most of these works apply multi-instance learning (MIL) loss due to its ability to learn discriminative representations under weak labels. 
Based on the MIL loss, CoLA \cite{CoLA} also introduced a snippet contrast loss to refine the hard snippet representation in feature space, which guides the network to perceive precise temporal boundaries. 
CO2-Net \cite{CO2N} devised a cross-modal consensus mechanism to capture inter-modality consistency and filter out task-irrelevant redundant information. 
CASE \cite{CASE} adopted unsupervised snippet clustering to explore the underlying structure among the snippets. 
SAL \cite{SAL} designed a multilevel semantic learning branch and an adaptive actionness learning branch to introduce second-order video semantics and learn class-agnostic action information, respectively. 
To model anomaly semantics, PE-MIL \cite{PE-MIL} and LPCF \cite{LPCF} adopted prompt-enhanced learning to detect various abnormal events while ensuring clear event boundaries.
Unlike these end-to-end MIL methods, P-MIL \cite{P-MIL} introduced an additional second stage to directly classify proposals based on temporal region of interest features. FuSTAL \cite{FUSTAL} incorporated an EMA-based distillation teacher model to enhance localization performance at the second stage.



However, these methods are not easily applicable to the wATFL task due to several differences. 
First, the modality is mismatched, as most approaches extract single-modality video features for analysis. 
Second, the tasks are different. wVAD is a detection task, whereas wTAL requires predicting timestamps of forgery proposals. 
In addition, the core of wTAL is the Temporal Class Activation Sequence (T-CAS), which is specifically designed for multi-class action recognition, whereas wATFL focuses on modeling temporal forgery probability using attention mechanisms. 
Therefore, the action proposals derived from T-CAS and attention scores contain excessive redundancy, requiring hyperparameter adjustments to achieve satisfactory performance.

\section{Methodology}

\subsection{Problem Definition} Assume we have $N$ untrimmed training audio $\{X_{n}\}_{n=1}^N$. The goal of wATFL is to train the model using utterance-level labels $Y_n\in \{0,1\}$, where $Y_n=1$ indicates a partial forgery audio and $Y_n=0$ otherwise.
During inference, the frame-level forgery scores $\{\hat{s}_t\}_{t=1}^{T}$ and a set of forgery proposals $\{(c, s, e)\}$ are predicted for each audio, where $T$ denote the number of frames, $c$, $s$ and $e$ represent the confidence score, the start timestamp and the end timestamp of each proposal, respectively.

\begin{figure*}[!t]
    \centering
    \includegraphics[width=0.9\linewidth]{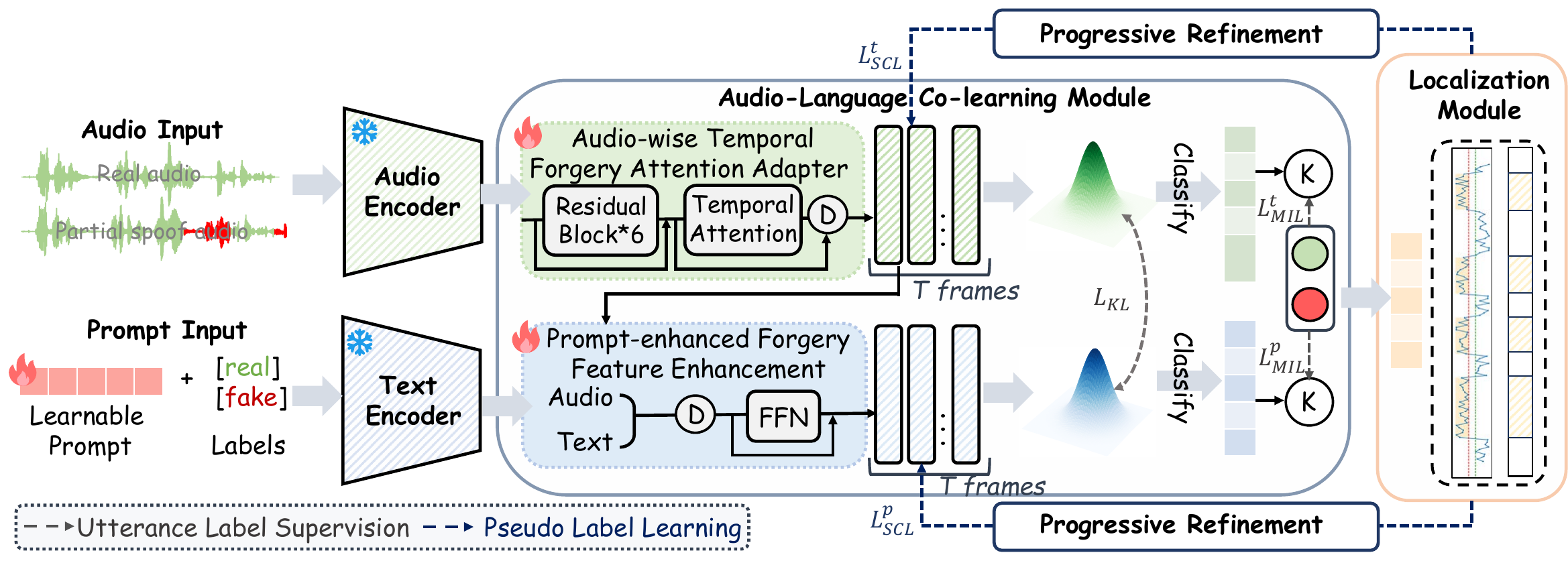}
\caption{The network structure of the proposed LOCO. It incorporates an A2LC module to capture forgery consensus features by aligning semantics from temporal and global perspectives, followed by a PRS to amplify the semantic distinction between real and fake content.}
    \label{fig:2}
\end{figure*}

\subsection{Overview} 
Due to the differences between weakly-supervised tasks, we specifically design a progressive audio-language co-learning network for the wATFL task, with the structure shown in Figure \ref{fig:2}. 
Based on the design principle of PFMs, this network aims to mine temporal forgery traces of semantic-driven modifications through co-learning and self-supervision manners to compensate for the lack of binary supervision.

First, the semantic information of partial forgery audio is deliberately modified through the pipeline shown in Figure \ref{fig:1}, such as insertion, deletion, and replacement operations. To this end, an audio-language co-learning module is devised to learn discriminative features by mining semantic inconsistency, as detailed in Section \ref{sec:1}.  Specifically, given an input audio, an automatic speech recognition (ASR) model is used to extract the audio features with semantic information. These features are then passed to a temporal forgery attention adapter to extract temporal-enhanced forgery features. 

Second, inspired by prompt-enhanced learning, the forgery-aware prompts are constructed using utterance-level annotations together with learnable prompts. Followed by a prompt-enhanced forgery feature adapter, semantic priors are incorporated into audio features dynamically to provide prompt-enhanced features with valuable contextual and semantic cues from a global view. 
In addition to using segmental MIL to guide these two discriminative features, a co-learning loss is introduced to ensure the alignment of semantics and features. 
 %

Finally, since MIL often focuses on segmental-level predictions, it ignores the relationships between instances within a segment (e.g., semantic information).
Thus, a progressive refinement strategy is designed to optimize forgery-aware features in a self-supervised manner by using semantic-based contrastive learning, as detailed in Section \ref{sec:2}. 

\subsection{Audio-Language Co-learning Module}
\label{sec:1}


The A2LC module aims to mine the contextual semantic inconsistencies introduced by partial forgery manipulations, thereby improving the model’s ability to detect forgery segments in partial forgery audio.
Specifically, the A2LC module consists of a TFA adapter and a PFF adapter to collaboratively capture forgery-aware features, which are optimized through a co-learning strategy, as depicted in the left part of Figure 2. 

\textbf{Temporal Forgery Attention Adapter}: Due to the principle behind semantical-driven manipulations, a pretrained audio-to-language (a.k.a, ASR) feature extractor, namely XLS-R-300M, is applied to extract initial audio features $F_a\in R^{T \times D}$ with content semantics, where $T$ denotes the number of frames in the audio and $D$ is the dimension of features. Subsequently, to model temporal forgery cues, a TFA adapter is applied to enhance features  through several residual blocks $\text{RB}(\cdot)$ with a temporal attention mechanism, and the latent forgery-aware features $F^t_a \in R^{T \times D'}$ are obtained as:

\begin{equation}
\begin{aligned}
        F^{rb}_a&= \text{RB}(\text{XLSR}(X), \alpha_{RB}) \\
        F^{t}_a &=  F^{rb}_a \odot \text{softmax}(F^{rb}_a)
\end{aligned}
\end{equation}
where $\odot$ denotes the element-wise multiplication, $\alpha_{RB}$ are learnable parameters, and $D'$ is the feature dimension. 
Subsequently, the frame-level scores $\hat{S}_t=FC(F^{t}_a, \alpha^t_{FC}) \in \mathbb{R}^{T \times 2}$ are obtained by applying a classifier layer $FC(\cdot)$, where $\alpha^t_{FC}$ are learnable parameters.

\textbf{Prompt-enhanced Forgery Feature Adapter}:
Based on prompt-enhanced learning \cite{PE-MIL}, forgery-aware prompts are introduced to provide context cues and semantic information for wATFL from a global view. Specifically, forgery-aware prompts are constructed by concatenating the category embedding for utterable-level class $c_i \in \{real,fake\} $, with the learnable embedding that consists of $l$ context tokens to form a complete sentence token, and thus, the input of text encoder for one class is presented as $\{e_1, ..., e_l,c_i\}$. 
The overall label prompt embedding $Prompt \in \mathbb{R}^{D'}$ is the $\text{CLS}$ token output of a pre-trained text encoder \cite{BERT}. With forgery-aware prompt and forgery features $F^t_a$ in hand, prompt-enhanced forgery features $F^p_a$ are generated by a feed-forward network (FFN) layer and a skip connection, presented as follows:
  
\begin{equation}
    F^p_a=FFN(F^t_a \odot Prompt)+Prompt
\end{equation}
Such an implementation allows prompt-enhanced features $F^p_a$ to capture the related forgery semantic context cues from audio.
Similarly, the frame-level predictions ${\hat{S}_p}$ are obtained by applying a classifier layer to prompt-enhanced latent features.

 
\textbf{Co-learning-based Objective Function}: This optimization process is designed to guide two forgery features to learn complementary information from each other based on a co-learning idea. 
Following the weakly-supervised methods, the MIL-based loss is first applied to optimize each forgery feature by selecting top-K frames with the highest confidence scores, which can be denoted as:
\begin{equation}
    \hat{Y}_v=\frac{1}{K} \sum_{i \in topK} {s}_{v,i}
\end{equation} 
where $K=50$, and $v \in \{t, p\}$ denotes the predictions based on forgery features $\{F^t_a, F^p_a\}$. 
Then, the frame-level MIL loss is computed using a parameter-free loss function $MSE_{pg}(\cdot)$, namely the P2sGrad-based mean squared error (P2sGrad-MSE) \cite{p2sgrad}, which alleviates the imbalance issue where true frames outnumber forgery frames, and can be presented as follows:
\begin{equation}
\begin{aligned}
    L_{MIL} &= L^t_{MIL}+L^p_{MIL}\\
    &=MSE_{pg}(\hat{Y}_t,Y)+MSE_{pg}(\hat{Y}_p,Y)
\end{aligned}
\end{equation}
where $Y$ is utterance-level label to indicate a partial spoof audio ($Y =1$) or a real audio ($Y =0$). 

Additionally, a Kullback-Leibler (KL) divergence is utilized to ensure the alignment of semantics and feature distributions. 
Given the KL-divergence is not bounded, i.e., $ D_{KL} \in \{0, \infty )\}$, taking the exponential of its negative value transforms the objective from maximization to minimization. Then, the transformed objective is bounded within $(0, 1]$, which is numerically advantageous. The loss for feature co-learning can be denoted as:
\begin{equation}
    L_{KL}=\exp(-D_{KL}[F^t_a || F^p_a])+\exp(-D_{KL}[F^p_a ||F^t_a ])
\end{equation}


\subsection{Forgery Localization Module}
\label{sec:2}

The forgery localization module aims to generate forgery proposals for each audio. 
Unlike the wTAL localization module, this localization module computes a temporal forgery-class activation sequence (T-FAS) specifically tailored for wATFL tasks. 
Specifically, instead of using background-suppressed operations, attention scores, and T-CAS to generate proposals, we apply the softmax function to the output scores $\{\hat{S}_t, \hat{S}_p\}$ along the second dimension. This operation ensures that the binary scores ('real' and 'fake') for each frame sum to 1, and can be denoted as follows:
\begin{equation}
\begin{aligned}
    A&=\lambda_a\text{softmax}(\hat{S}_t,dim=1)\\
    &+(1-\lambda_a) \text{softmax}(\hat{S}_p,dim=1)
\end{aligned}
\end{equation}
where $\lambda_a=0.9$ and the second class represents the T-FAS, denoted as $A=\{a_t\}_{t=1}^T$, as no proposals are needed for real segments. Each score in $A$ indicates the probability of forgery for each frame. %
Instead of using a multiple thresholds strategy, we select forgery-aware threshold $\theta_f=0.5$ on the T-FAS to reduce redundancy proposals. Then, the forgery proposals $P=\{(c_m, s_m,e_m)\}_{m=1}^M$ are generated by concatenating consecutive forgery frames, where the confidence score is computed as:
\begin{equation}
    c_m=\frac{1}{e_m-s_m}\sum_{i=s_m}^{e_m}a_{i}
\end{equation}


\subsection{Progressive Refinement Strategy}




Given that MIL focuses on segmental-level predictions, it is insufficient to model the relationships between audio frames (e.g., semantic information). 
The fundamental reason is the lack of fine-grained annotations for forgery feature optimization. Thus, based on a self-supervised idea, a progressive refinement strategy is designed to obtain frame-level labels for free. To ensure the accuracy of the pseudo frame-level labels and prevent the introduction of additional training noise, a two-stage process is applied for training LOCO.

In the first training phase, the base model is trained for 500k steps, with the aim of capturing semantic inconsistencies for forgery cues and the objective function is denoted as:

\begin{equation}
L_{first} = L_{MIL} + \lambda_1 L_{KL}
\end{equation}
where $\lambda_1$ is set to 0.1 to balance the training loss.

Then, in the second phase, pseudo frame-level labels are calculated using the start and end timestamps of proposals $P=\{{c_m,s_m,e_m}\}_{m=1}^M$ predicted from the base model as:
\begin{equation}
\bar{y}_t = 
\begin{cases} 
1, & \text{if } 
s_m \cdot {fps} \leq t \leq e_m \cdot {fps}, \\
0, & \text{otherwise}
\end{cases}
\end{equation}
where $\bar{y}_t$ is the $t$-th frame in pseudo labels, $M$ denotes the number of predicted proposals, and $fps=50$ is the frames per second under a temporal resolution of 20 ms.

\begin{table*}[!t]
  \centering
  \caption{Frame-level evaluation results on three different datasets in terms of frame-level EER(\%), ACC(\%), and AUC(\%). Performance comparison with state-of-the-art full-supervised and weak-supervised detection methods, where $^\dagger$ means only the utterance-level loss is adopted.}
    \resizebox{0.83\textwidth}{!}{%
    \begin{tabular}{ccccccccccc}
    \toprule
    \multirow{2}{*}{Supervision} & \multirow{2}{*}{Method} & \multicolumn{3}{c}{HAD} & \multicolumn{3}{c}{LAV-DF} & \multicolumn{3}{c}{AV-Deepfake-1M} \\
\cmidrule{3-11}          &       & EER ($\downarrow$)  & AUC ($\uparrow$)   & ACC ($\uparrow$)  & EER ($\downarrow$)  & AUC ($\uparrow$)  & ACC ($\uparrow$)  & EER ($\downarrow$)  & AUC ($\uparrow$)  & ACC ($\uparrow$) \\
    \midrule
    \multirow{3}[2]{*}{Fully} & PSDL  & 0.35  & 99.98 & 99.78 & 1.07  & 99.88 & 99.14 & 0.48  & 99.97 & 99.52 \\
          & IFBDN & 0.18  & 99.97 & 99.89 & 0.82  & 99.92 & 98.93 & 0.71  & 99.92 & 99.29 \\
          & FDN   & 0.08  & 99.96 & 99.95 & 0.82  & 99.89 & 99.21 & 0.24  & 99.98 & 99.76 \\
\cmidrule{1-11}    \multirow{6}[2]{*}{Weakly} & IFBDN$^\dagger$ & 10.28 & 94.59 & 89.72 & 38.42 & 64.01 & 61.58 & 18.85 & 89.70 & 53.31 \\
          & FDN$^\dagger$  & 8.47 & 95.88 & 91.53 & 20.65 & 86.74 & 79.35 & 26.52 & 79.99 & 73.47 \\
          \cdashline{2-11} & CoLA  & 10.79 & 94.43 & 89.21 & 29.01 & 77.39 & 71.00 & 46.76 & 54.91 & 53.24 \\
          & CASE  & 9.59  & 94.84 & 90.41 & 27.33 & 79.41 & 72.67 & 27.21 & 80.67 & 72.79 \\
          & FuSTAL & 9.03  & 95.29 & 90.97 & 28.55 & 78.03 & 71.49 & 47.36 & 54.11 & 52.64 \\
          & \cellcolor[rgb]{ .906,  .902,  .902}\textbf{Ours} & \cellcolor[rgb]{ .906,  .902,  .902}\textbf{4.56} & \cellcolor[rgb]{ .906,  .902,  .902}\textbf{97.51} & \cellcolor[rgb]{ .906,  .902,  .902}\textbf{95.93} & \cellcolor[rgb]{ .906,  .902,  .902}\textbf{3.39} & \cellcolor[rgb]{ .906,  .902,  .902}\textbf{98.71} & \cellcolor[rgb]{ .906,  .902,  .902}\textbf{99.09} & \cellcolor[rgb]{ .906,  .902,  .902}\textbf{6.46} & \cellcolor[rgb]{ .906,  .902,  .902}\textbf{97.32} & \cellcolor[rgb]{ .906,  .902,  .902}\textbf{95.93} \\
    \bottomrule
    \end{tabular}%
    }%
  \label{tab:det}%
\end{table*}%

With these fine-grained labels, a semantic-based contrastive learning loss is designed to pull pairs of semantically similar audio frames in the feature space while pushing apart pairs with differing semantics, which can be presented as:

\begin{equation}
\begin{aligned}
    L_{SCL}&=\frac{1}{R} \sum^R_{r=1}I_r(1-SIM(f_r,f_r^+))^2\\ 
    &+(1-I_r)\max(0,SIM(f_r,f_r^-))^2
\end{aligned}
\end{equation}
where $SIM(\cdot)$ is the cosine similarity between two frames, $R$ denotes the number of frame pairs, $I_r=1$ (or $I_r=0$) indicates a similar (or dissimilar) frame in the $r$-th pair, $f_r$ and $f_r^+$ (or $f_r^-$) present the features of reference frame and similar (or dissimilar) frame, respectively.

Then, the objective function of the second stage is computed as:

\begin{equation}
L_{second} = L_{MIL} + \lambda_1 L_{KL} + \lambda_2 L_{SCL}
\end{equation}
where $\lambda_1=0.1$ and $\lambda_2=0.01$ are hyperparameters to balance the training loss. Finally, the pseudo frame-level labels are iteratively refined using predictions from the previous iteration, progressively improving the localization performance.



 




\label{sec:3}

\section{Experimental results}
\subsection{Experimental Settings}

\textbf{Dataset.}
We evaluate our method on three datasets, including LAV-DF \cite{BATFD}, HAD \cite{HAD}, and AV-Deepfake-1M \cite{AVDF1M}, where LAV-DF and HAD adopt a rule-based replacement to modify audio while AV-Deepfake-1M is LLM-driven manipulation. 
Due to the lack of testing labels in the audio modality of AV-Deepfake-1M, we combine the training and validation sets to construct a new subset, ensuring non-overlapping speakers according to the original division strategy. Finally, we select 83,517, 30,212, and 20,417 audio for training, validation, and testing, respectively, according to the number and split ratio of other datasets.

\textbf{Evaluation Metrics.}
We evaluate the proposed LOCO in terms of detection and localization performance. 
Following existing works, we use frame-level equal error rate (EER), accuracy (ACC), and area under the curve (AUC) to evaluate the detection task. For the localization task, we adopt average precision (AP) at the temporal intersection over union (IoU) thresholds $[0.1:0.9:0.1]$, average recall (AR) with different average number of proposals (AN) $\{2,5,10,20\}$.

\textbf{Compared Methods.}
We select several fully-supervised start-of-the-art countermeasures, including detection methods (e.g., PSDL\cite{PSDL}, IFBDN \cite{IFBDN}, FDN \cite{CFPRF}) and localization methods (e.g., BA-TFD\cite{BATFD}, BA-TFD+ \cite{BATFD2}, UMMAFormer \cite{UMMAF}, CFPRF \cite{CFPRF}). 

Given the lack of wATFL methods, we adapt several wTAL methods (e.g., CoLA \cite{CoLA}, CASE \cite{CASE}, FuSTAL \cite{FUSTAL},  SAL \cite{SAL}) for comparison with two key modifications. First, we replace their original input features with the XLS-R features used in our method to ensure a fair comparison. Second, we apply a softmax function to T-CAS and change the dimension to 2 to suit the wATFL task, while their original multi-threshold strategy for proposal generation is retained. 
Additionally, we remove fine-grained supervision losses from the fully-supervised detection method and retain only utterance-level losses to assess their detection performance under weak supervision scenarios.

\subsection{Implement Detail}
All experiments are conducted on a single GeForce RTX 3090 GPU. 
In terms of the model input, we preprocess each audio in a single-channel format with a 16-kHz sampling rate, then pre-trained XLS-R-300M \cite{XLSR} and bert-base-uncased  \cite{BERT} are adopted to extract initial audio and textual features, respectively. 
For model training, we utilize the Adam optimizer with a batch size of 2 and a learning rate of $10^{-6}$. The base model is trained for 300k steps on the HAD and LAV-DF datasets and 900k steps on the AV-Deepfake-1M dataset.
Subsequently, the base model is trained for 15 epochs in the second stage and selected using the best results on the validation portion for testing.

\begin{table*}[!t]
  \centering
  \caption{Localization evaluation results on the evaluation datasets in terms of AP@IoU(\%), mAP(\%) and AR@AN(\%). Performance comparison with state-of-the-art fully-supervised and weakly-supervised localization methods.}
    \resizebox{0.98\textwidth}{!}{
 \begin{tabular}{ccccccccccccccccc}
    \toprule
    \multirow{2}{*}{Dataset} & \multirow{2}{*}{Method} & \multirow{2}{*}{Super.} & \multicolumn{9}{c}{AP@IoU($\uparrow$)}                                        & \multirow{2}{*}{mAP($\uparrow$)} & \multicolumn{4}{c}{AR@AN ($\uparrow$)} \\
\cmidrule{4-12}\cmidrule{14-17}          &       &       & 0.1   & 0.2   & 0.3   & 0.4   & 0.5   & 0.6   & 0.7   & 0.8   & 0.9   &       & 2     & 5     & 10    & 20 \\
    \midrule
    \multirow{9}{*}{\rotatebox{90}{HAD}} & BA-TFD & Fully & 92.06  & 91.89  & 90.81  & 88.00  & 79.86  & 64.90  & 46.03  & 29.49  & 5.55  & 65.40  & 71.59  & 73.31  & 74.80  & 76.72  \\
          & BA-TFD+ & Fully & 93.66  & 93.59  & 93.01  & 91.39  & 88.26  & 83.22  & 76.11  & 64.10  & 37.83  & 80.13  & 84.06  & 84.61  & 84.87  & 85.34  \\
          & UMMAF & Fully & 99.98  & 99.98  & 99.98  & 99.98  & 99.98  & 99.95  & 99.88  & 99.74  & 98.01  & 99.72  & 99.75  & 99.81  & 99.81  & 99.81  \\
          & CFPRF & Fully & 99.77  & 99.79  & 99.79  & 99.80  & 99.77  & 99.73  & 99.66  & 99.48  & 99.21  & 99.67  & 99.73  & 99.73  & 99.73  & 99.73  \\
\cmidrule{2-17}          & CoLA  & Weakly & 64.50  & 61.68  & 59.63  & 58.01  & 55.46  & 50.24  & 40.00  & 21.27  & 2.91  & 45.97  & 55.34  & 59.25  & 60.82  & 61.00  \\
          & CASE  & Weakly & 81.68  & 65.60  & 56.05  & 49.54  & 46.17  & 43.12  & 39.58  & 32.33  & \textbf{9.43 } & 47.06  & 55.41  & 61.17  & 61.54  & 61.57  \\
          & FuSTAL & Weakly & 83.33  & 82.35  & 81.15  & 78.46  & 72.59  & 60.56  & 42.25  & 19.09  & 2.81  & 58.07  & 64.69  & 68.77  & 70.74  & 71.09  \\
          & SAL   & Weakly & 69.36  & 66.39  & 64.99  & 63.94  & 62.58  & 59.01  & 49.85  & 27.52  & 3.28  & 51.88  & 61.81  & 74.01  & 77.35  & 77.47  \\
          & \cellcolor[rgb]{ .906,  .902,  .902} \textbf{Ours} & \cellcolor[rgb]{ .906,  .902,  .902}\textbf{Weakly} & \cellcolor[rgb]{ .906,  .902,  .902}\textbf{97.44} & \cellcolor[rgb]{ .906,  .902,  .902}\textbf{97.41} & \cellcolor[rgb]{ .906,  .902,  .902}\textbf{96.73} & \cellcolor[rgb]{ .906,  .902,  .902}\textbf{95.33} & \cellcolor[rgb]{ .906,  .902,  .902}\textbf{93.17} & \cellcolor[rgb]{ .906,  .902,  .902}\textbf{89.22} & \cellcolor[rgb]{ .906,  .902,  .902}\textbf{77.33} & \cellcolor[rgb]{ .906,  .902,  .902}\textbf{39.48}  &  \cellcolor[rgb]{ .906,  .902,  .902}5.51 & \cellcolor[rgb]{ .906,  .902,  .902}\textbf{76.85} & \cellcolor[rgb]{ .906,  .902,  .902}\textbf{82.02} & \cellcolor[rgb]{ .906,  .902,  .902}\textbf{82.43} & \cellcolor[rgb]{ .906,  .902,  .902}\textbf{82.44} & \cellcolor[rgb]{ .906,  .902,  .902}\textbf{82.44} \\
    \midrule
    \multirow{9}{*}{\rotatebox{90}{LAV-DF}}& BA-TFD & Fully & 60.28  & 59.68  & 59.01  & 57.83  & 53.53  & 39.42  & 19.23  & 5.26  & 0.36  & 39.40  & 52.77  & 56.24  & 59.83  & 65.84  \\
          & BA-TFD+ & Fully & 86.27  & 86.07  & 85.76  & 85.30  & 83.78  & 77.34  & 63.54  & 37.40  & 6.13  & 67.95  & 72.21  & 75.00  & 76.36  & 77.65  \\
          & UMMAF & Fully & 98.20  & 98.02  & 97.82  & 97.56  & 97.29  & 96.86  & 96.11  & 95.07  & 89.91  & 96.32  & 95.22  & 98.36  & 98.79  & 99.08  \\
          & CFPRF & Fully & 97.71  & 96.93  & 96.11  & 95.19  & 94.52  & 94.08  & 93.63  & 93.30  & 91.65  & 94.79  & 95.29  & 95.30  & 95.30  & 95.30  \\
\cmidrule{2-17}          & CoLA  & Weakly & 51.60  & 45.80  & 40.42  & 36.54  & 33.60  & 31.27  & 28.18  & 21.67  & 4.65  & 32.64  & 55.02  & 62.92  & 67.13  & 68.86  \\
          & CASE  & Weakly & 58.63  & 50.44  & 44.90  & 42.03  & 40.20  & 38.87  & 37.56  & 34.96  & 21.97  & 41.06  & 65.31  & 71.88  & 72.06  & 72.15  \\
          & FuSTAL & Weakly & 61.69  & 59.92  & 57.68  & 55.27  & 51.70  & 47.07  & 39.26  & 24.34  & 2.94  & 44.43  & 59.52  & 65.85  & 68.72  & 69.75  \\
          & SAL   & Weakly & 75.12  & 69.27  & 64.29  & 60.41  & 57.06  & 54.16  & 51.80  & 47.75  & 26.14  & 56.22  & 68.14  & 73.42  & 73.72  & 73.75  \\
          & \cellcolor[rgb]{ .906,  .902,  .902} \cellcolor[rgb]{ .906,  .902,  .902}\textbf{Ours} & \cellcolor[rgb]{ .906,  .902,  .902}\textbf{Weakly} & \cellcolor[rgb]{ .906,  .902,  .902}\textbf{95.66} & \cellcolor[rgb]{ .906,  .902,  .902}\textbf{94.02} &  \cellcolor[rgb]{ .906,  .902,  .902}\textbf{92.23} & \cellcolor[rgb]{ .906,  .902,  .902}\textbf{89.71} & \cellcolor[rgb]{ .906,  .902,  .902}\textbf{87.78} & \cellcolor[rgb]{ .906,  .902,  .902}\textbf{85.33} & \cellcolor[rgb]{ .906,  .902,  .902}\textbf{82.03} & \cellcolor[rgb]{ .906,  .902,  .902}\textbf{75.73} & \cellcolor[rgb]{ .906,  .902,  .902}\textbf{41.37} & \cellcolor[rgb]{ .906,  .902,  .902}\textbf{82.65} & \cellcolor[rgb]{ .906,  .902,  .902}\textbf{88.81} & \cellcolor[rgb]{ .906,  .902,  .902}\textbf{89.49} & \cellcolor[rgb]{ .906,  .902,  .902}\textbf{89.49} & \cellcolor[rgb]{ .906,  .902,  .902}\textbf{89.49} \\
    \midrule
    \multirow{8}{*}{\rotatebox{90}{AV-Deepfake-1M}}& BA-TFD & Fully & 44.61  & 43.58  & 39.91  & 28.57  & 15.49  & 6.99  & 1.92  & \diagbox{}{}  & \diagbox{}{}  & 27.78  & 33.71  & 40.35  & 44.75  & 48.84  \\
          & UMMAF & Fully & 96.77 & 96.76 & 96.76 & 96.73 & 96.67 & 96.57 & 96.33 & \diagbox{}{} & \diagbox{}{} & 96.66 & 95.97 & 98.71 & 98.98 & 99.08 \\
          & CFPRF & Fully & 94.11  & 94.06  & 93.96  & 93.74  & 93.06  & 92.49  & 91.77  & \diagbox{}{}  & \diagbox{}{}  & 93.31  & 92.97  & 93.47  & 93.47  & 93.47  \\
\cmidrule{2-17}          & CoLA  & Weakly & 5.07  & 2.49  & 1.37  & 0.38  & 0.11  & 0.03  & 0.01  & \diagbox{}{}  & \diagbox{}{}  & 1.35  & 2.77  & 6.12  & 11.44  & 17.14  \\
          & CASE  & Weakly & 14.79  & 8.43  & 5.58  & 3.97  & 2.77  & 1.55  & \textbf{0.53}  & \diagbox{}{}  & \diagbox{}{}  & 5.37  & 13.64  & 24.08  & 30.18  & 32.54  \\
          & FuSTAL & Weakly & 7.77  & 3.48  & 1.41  & 0.53  & 0.18  & 0.06  & 0.01  & \diagbox{}{}  & \diagbox{}{}  & 1.92  & 4.68  & 9.46  & 16.24  & 24.66  \\
          & SAL   & Weakly & 15.67  & 10.42  & 6.99  & 4.58  & 2.75  & 1.13  & 0.23  & \diagbox{}{}  & \diagbox{}{}   & 5.97  & 17.27  & 28.48  & 22.64  & 15.83  \\
           & \cellcolor[rgb]{ .906,  .902,  .902} \textbf{Ours} & \cellcolor[rgb]{ .906,  .902,  .902}\textbf{Weakly} & \cellcolor[rgb]{ .906,  .902,  .902}\textbf{42.88} & \cellcolor[rgb]{ .906,  .902,  .902}\textbf{40.31} & \cellcolor[rgb]{ .906,  .902,  .902}\textbf{32.99} & \cellcolor[rgb]{ .906,  .902,  .902}\textbf{25.24}  & \cellcolor[rgb]{ .906,  .902,  .902}\textbf{15.50}  & \cellcolor[rgb]{ .906,  .902,  .902}\textbf{3.82} &  \cellcolor[rgb]{ .906,  .902,  .902}0.26  & \cellcolor[rgb]{ .906,  .902,  .902}\diagbox{}{}  & \cellcolor[rgb]{ .906,  .902,  .902}\diagbox{}{} & \cellcolor[rgb]{ .906,  .902,  .902}\textbf{18.89} & \cellcolor[rgb]{ .906,  .902,  .902}\textbf{29.55} & \cellcolor[rgb]{ .906,  .902,  .902}\textbf{46.65} & \cellcolor[rgb]{ .906,  .902,  .902}\textbf{49.44} & \cellcolor[rgb]{ .906,  .902,  .902}\textbf{49.70}  \\
    \bottomrule
    \end{tabular}%
    }
  \label{tab:loc}%
\end{table*}%
\vspace{-0.2em}

\subsection{Comparison with State-of-the-art Methods}
\textbf{Frame-level Forgery Detection Results.} 
Table \ref{tab:det} presents the frame-level detection results of the proposed LOCO compared with weakly-supervised and fully-supervised methods on three datasets. 
The results indicate that LOCO outperforms other state-of-the-art weakly-supervised methods, reducing EER values of 3.91\%,  17.26\% and 12.39\% on the HAD, LAV-DF, and AV-Deepfake-1M datasets, respectively, compared to the second-best method.
Additionally, we observe that while fully-supervised methods achieve significant performance, their performance drops drastically when only utterance-level annotations are used for training.
For example, the best-performing model, denoted as FDN $\rightarrow$ FDN$^{\dagger}$, increases EER value from 0.08\% to 8.47\% on the HAD dataset.
This may be due to these methods relying on fine-grained label-based losses to guide complex modules for feature learning, making them unsuitable for direct adaptation to weakly-supervised settings.
These results demonstrate the effectiveness of our proposed LOCO for the detection task.





\textbf{Temporal Forgery Localization Results.}
Table \ref{tab:loc} shows the temporal forgery localization comparison results on three datasets. From the results, we can see that the proposed LOCO outperforms the prior state-of-the-art weakly-supervised methods. 
Specifically, LOCO surpasses the previous best performance by 18.78\% and 26.43\% in terms of the mAP on the HAD and LAV-DF dataset, respectively.
Additionally, our method achieves 18.89\% mAP, representing a 12.92\% improvement over the second-best method. This may be due to AV-Deepfake-1M containing more long-duration samples, which pose significant challenges for weakly-supervised performance, especially under high IoU thresholds.
Moreover, even when compared to certain fully-supervised methods (e.g. BA-TFD and BA-TFD+), our model can achieve comparable results at low IoU ($<0.5$) thresholds.
The clear localization performance superiority demonstrates the effectiveness of the proposed LOCO again.

\subsection{Ablation Studies}

\begin{table}[!t]
  \centering
  \caption{Ablation study of LOCO with different components in terms of mAP(\%) on the HAD dataset}
    \resizebox{0.29\textwidth}{!}{\begin{tabular}{ccc}
    \toprule
    Structure & mAP($\uparrow$)   & $\Delta$ \\
    \midrule
    \rowcolor[rgb]{ .949,  .949,  .949} Baseline=LOCO & 76.85  & \diagbox{}{} \\
    \midrule
    freeze($E_A$)$\rightarrow$train($E_A$) & 8.95    & -67.9 \\
    w/o. TFA & 70.03 & -6.82 \\
    w/o. PFF & 74.59  & -2.26 \\
    w/o. PRS & 73.81    & -3.04 \\
    \bottomrule
    \end{tabular}}
  \label{tab:com}
\end{table}

\begin{table}[!t]
  \centering
  \caption{Ablation study of LOCO with different loss settings in terms of mAP(\%) on the HAD dataset}
    \resizebox{0.42\textwidth}{!}{
    \begin{tabular}{cccc}
    \toprule
    Structure & Loss Settings & mAP($\uparrow$)   & $\Delta$ \\
    \midrule
    \rowcolor[rgb]{ .949,  .949,  .949} Baseline=LOCO & $l_{MIL}+l_{KL}+l_{SCL}$ &  76.85 & \diagbox{}{} \\
    \midrule
    \multirow{3}[2]{*}{w/o. PRS} 
    & $l^t_{MIL}(BCE)$ & 62.12  & -14.73 \\
    & $l^t_{MIL}(P2sGrad)$ & 68.79  & -8.06 \\
          & $l^t_{MIL}+l^{p}_{MIL}$ & 71.65  & -5.20 \\
          & $l_{MIL}+l_{KL}$ & 73.81    & -3.04 \\
    \bottomrule
    \end{tabular}%
    }
  \label{tab:loss}%
\end{table}
\begin{table}[!t]
  \centering
    \caption{Ablation study of LOCO with different localization thresholds in terms of mAP(\%)  on the LAV-DF dataset}
    \resizebox{0.35\textwidth}{!}{
    \begin{tabular}{cccccc}
    \toprule
    $\theta$ & 0.5 &  0.6   & 0.7   & 0.8   & 0.9 \\
    \midrule
    mAP($\uparrow$) &82.65 & 82.34 & 81.92 & 81.44 & 80.57 \\
    \bottomrule
    \end{tabular}%
    }
  \label{tab:th}%
\end{table}%

\textbf{Impact of different components.} As shown in Table \ref{tab:com}, we investigate the contribution of the core components of the proposed LOCO. Specifically, we first conduct experiments with two training strategies for the TFA adapter (e.g., freezing and training the XLS-R feature extractor) to show whether the features can effectively adapt. And we observe that pretrained features are more suitable for wATFL because these features are derived from the large-scale and diverse audio corpus and have already captured rich semantic representations. In contrast, unfreezing and training the features may lead to disrupting the generalizable pretrained representations based on limited wATFL data. Additionally, replacing the TFA adapter with an FFN results in a 6.82\% mAP decrease, highlighting the importance of audio features enhanced with temporal attention for effective localization.
Moreover, we observe a 2.26\% mAP decrease when the PFF adapter is removed, indicating that forgery-aware prompts are crucial for providing contextual semantic features.
These results further demonstrate the effectiveness of the co-learning strategy in the A2LC module. 
Finally, benefiting from the learning of contrastive semantic features, the progressive refinement strategy improves the mAP values by 3.04\%, respectively. 
All in all, each component of the LOCO serves a significant purpose and is thoughtfully designed.

\textbf{Impact of loss functions.} As shown in Table \ref{tab:loss}, we explore the contribution of each loss function in guiding forgery-aware feature learning. The MIL loss, serving as the foundational loss, enables temporal forgery features to achieve a localization performance of 68.79\% mAP on the HAD dataset. 
When the P2sGrad-based MSE loss within MIL is replaced with the Binary CrossEntropy (BCE) loss, a reduction of 6.67\% in mAP value is observed. 
This demonstrates that the P2sGrad-based loss is more suitable for the ATFL task, as it adaptively mitigates the bias of misclassifying fake frames as real without requiring the manual tuning of hyperparameters. 
With the addition of prompt-enhanced forgery features guided by the MIL loss, the localization performance improves by 2.86\%. 
Furthermore, the inclusion of the KL-divergence loss to guide the co-learning of forgery features yields an additional 2.16\% improvement in mAP performance. 
Finally, incorporating the semantic contrastive learning loss to enhance the distinction between different semantic content results in a further mAP increase of 3.04\%.

\textbf{Impact of localization thresholds}
Table \ref{tab:th} shows the results at different forgery thresholds for generating localization proposals. As the threshold increases, the localization precision for forgery segments gradually decreases. The best mAP is achieved at a threshold of 0.5.

\begin{figure}[!t]
    \centering
    \includegraphics[width=0.83\linewidth]{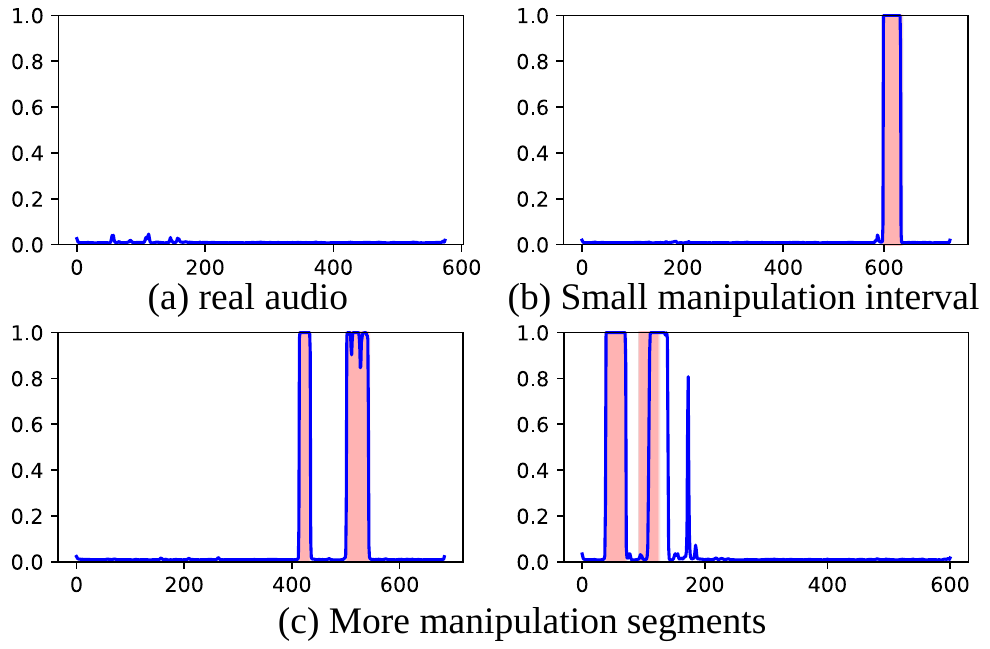}
    \caption{Qualitative prediction results based on LOCO's prediction scores where red region indicates forgery segments.}
    \label{fig:3}
\end{figure}

\begin{figure}[!t]
    \centering
    \includegraphics[width=1\linewidth]{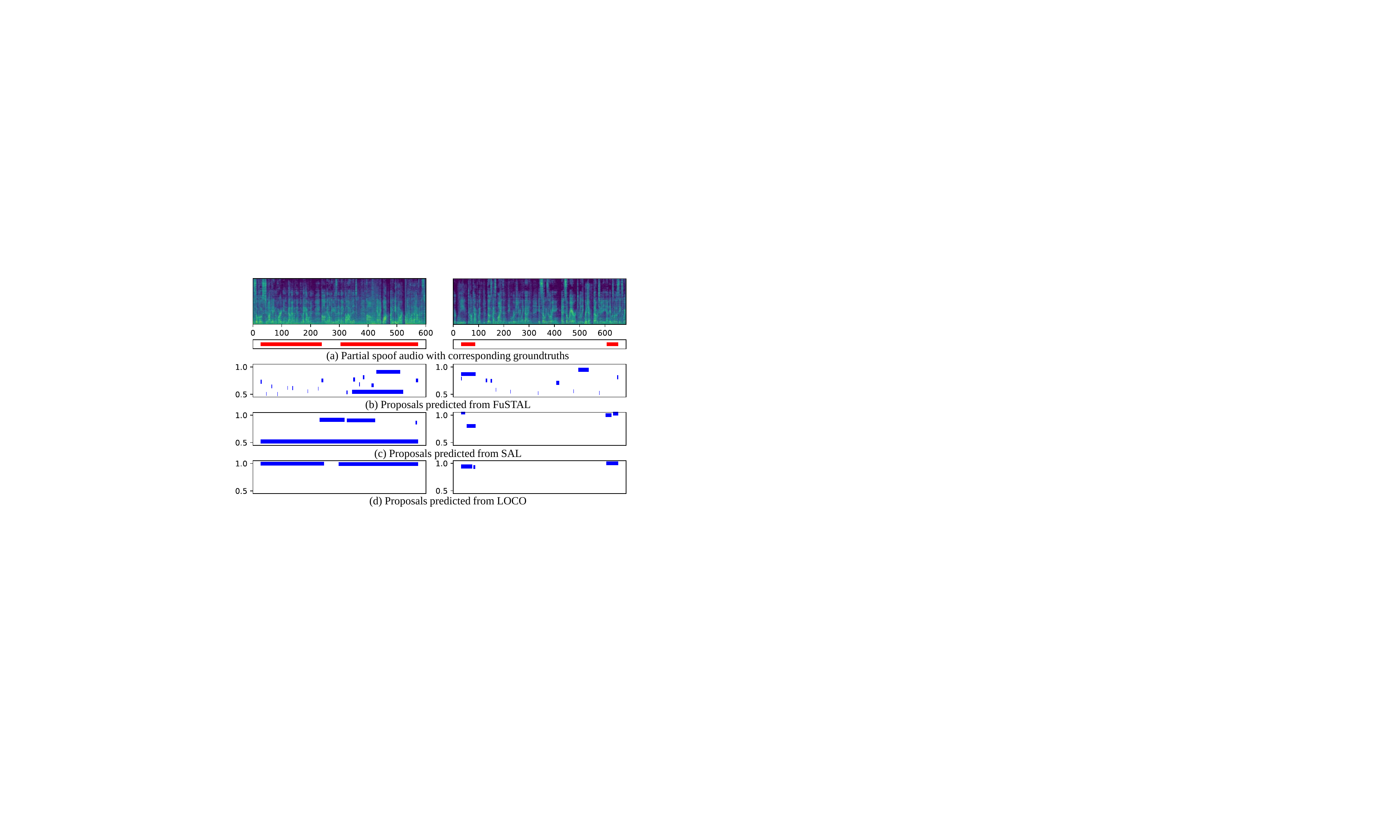}
    \caption{Qualitative localization comparison with ground-truth annotations and comparison methods in terms of predicted proposals. }
    \label{fig:4}
\end{figure}

\subsection{Qualitative Results}
To gain more insight, the frame-level prediction scores predicted by our method are visualized in Figure \ref{fig:3}. As demonstrated in Figure \ref{fig:3} (a),  our method mitigates false alarms in normal audio. 
Figure \ref{fig:3}(b) and Figure \ref{fig:3}(c) exemplify the proficiency of our method in predicting precise forgery scores for long-term partial forgery audio with multi-segment and subtle manipulation segments. The ability to detect frame-level forgery content demonstrates the effectiveness of forgery-aware features based on semantic inconsistencies in capturing partial spoof manipulations.

Then, to intuitively demonstrate the localization performance compared to other SOTA methods, we show qualitative localization results in Figure \ref{fig:4}. It is evident that LOCO produces more continuous and accurate forgery proposals, even in cases of multi-segment forgery (left part of Figure \ref{fig:4}) and subtle forgery spanning long durations (right part of Figure \ref{fig:4}). In contrast, the compared methods tend to generate multiple, disconnected, shorter forgery proposals. 

\begin{figure}[!t]
    \centering
    \includegraphics[width=0.85\linewidth]{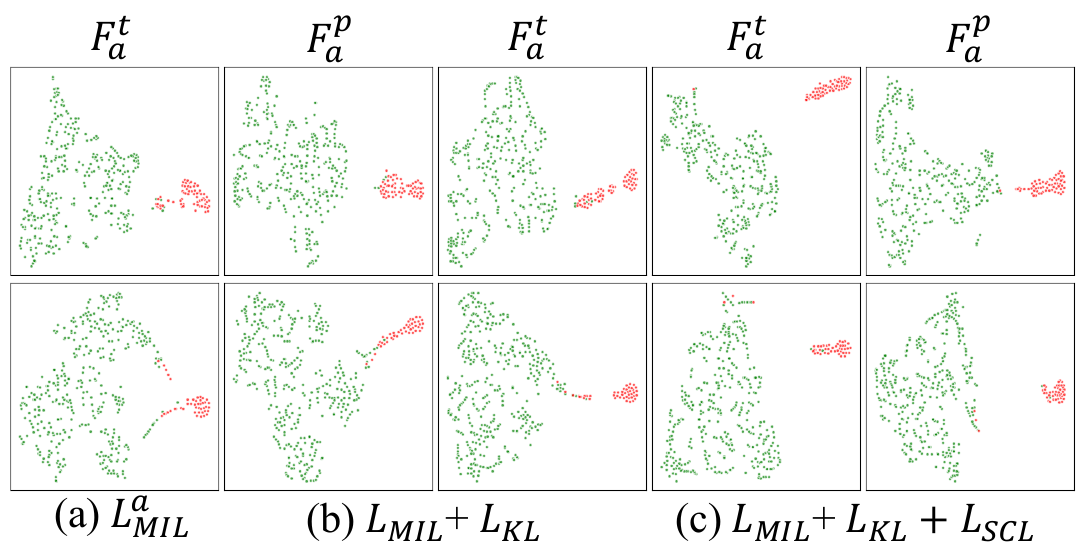}
    \caption{t-SNE visualization of latent features with different loss functions. Green and red dots denote real and forgery frames.} 
    \label{fig:5}
\end{figure}

Finally, to demonstrate the effect of co-learning loss and semantic contrastive learning, we employ t-SNE \cite{tsne} to visualize the latent features from the intermediate layer. 
In Figure \ref{fig:5}(a), we find that real and forgery frame-level features tend to be mixed at the boundaries when only audio features used. After incorporating prompt features for co-learning, this issue is alleviated, however, some forgery features remain mingled with the real ones, as shown in Figure \ref{fig:5}(b).
With the inclusion of the SCL loss, a more distinct separation emerges between real and forgery frames, as depicted in Figure \ref{fig:5}(c). These findings demonstrate that co-learning and the SCL loss effectively enhance the distinction between real and forgery frame-level features.

\vspace{-0.5em}
\section{Conclusion}
In this work, we present a novel method LOCO for wATFL, which aims to capture semantical inconsistencies and amplify the semantical difference between real and forgery content. Thus, an audio-language co-learning module is introduced to ensure the alignment of semantics and features from temporal and global perspectives. Then, a localization module is applied to generate proposals for each audio, where pseudo frame-level labels are used in the next stage of the progressive refinement strategy. 
In such a self-supervised manner, forgery-aware features are enriched to generate a clear forgery boundary.
Extensive experiments show our method achieves SOTA performance on three public benchmarks.
In the future, how to improve localization under a more precise IoU threshold (e.g., excessive 0.9) and long-duration samples, is a problem worthy of long-standing research.




\section*{Acknowledgments}
This work is supported by the National Natural Science Foundation of China (No.62441237, No.62261160653, No.62172435), the Guangdong Provincial Key Laboratory of Information Security Technology (No. 2023B1212060026), and Henan Province Innovation Scientists and Technicians Troop Construction Projects of China (No.254000510007).


\bibliographystyle{named}
\bibliography{ijcai25}

\end{document}